\documentclass[12pt]{article}

\usepackage{epsf}

\newcommand{\be}{\begin{equation}}
\newcommand{\ee}{\end{equation}}
\newcommand{\bea}{\begin{eqnarray}}
\newcommand{\eea}{\end{eqnarray}}
\newcommand{\beq}{\begin{equation}}
\newcommand{\eeq}{\end{equation}}
\newcommand{\ba}{\begin{array}}
\newcommand{\ea}{\end{array}}
\newcommand{\beqa}{\begin{eqnarray}}
\newcommand{\eeqa}{\end{eqnarray}}

\newcommand{\cL}{{\cal L}}
\newcommand{\cH}{{\cal H}}

\newcommand{\cO}{{\cal O}}
\newcommand{\da}{^\dagger}
\newcommand{\no}{\nonumber}
\newcommand{\nnu}{\nonumber}
\newcommand{\lsim}{\stackrel{<}{_\sim}}
\newcommand{\gsim}{\stackrel{>}{_\sim}}

\newcommand{\dbar}{{\overline d}}
\newcommand{\ubar}{{\overline u}}
\newcommand{\lbar}{{\overline \ell}}

\def\npb#1#2#3{    {\it Nucl. Phys. }{\bf B #1} (#2) #3}
\def\plb#1#2#3{    {\it Phys. Lett. }{\bf B #1} (#2) #3}

\def\prd#1#2#3{    {\it Phys. Rev. }{\bf D #1} (#2) #3}

\def\prl#1#2#3{    {\it Phys. Rev. Lett. }{\bf #1} (#2) #3}

\def\rmp#1#2#3{    {\it Rev. Mod. Phys. }{\bf #1} (#2) #3}

\def\jhep#1#2#3{   {\it JHEP  }{\bf #1} (#2) #3}

\begin{document}
%%%%%%%%%%%%%%%%%%%%%%%%%%%%%%%%%%%%%%%%%%%%%%%%%%%%%%%%%%%%%%%%%%%%%%% 

\thispagestyle{empty}
\begin{flushright}
CERN--TH/2001--266 \\
hep-ph/0110121 \\
October 2001 
\end{flushright}
\vskip 3.0 true cm 

\begin{center}
{\Large\bf Scalar flavour-changing neutral currents  \\ [5 pt]
           in the large-$\tan\beta$ limit} 
 \\ [25 pt]
{\sc {\sc Gino Isidori${}^{a,b}$ and Alessandra Retico${}^{a,c}$ } 
 \\ [25 pt]
{\sl ${}^a$Theory Division, CERN, CH-1211 Geneva 23, Switzerland} \\ [5 pt]
{\sl ${}^b$INFN, Laboratori Nazionali di Frascati, Via E. Fermi 40, 
           I-00044 Frascati, Italy} \\ [5 pt] 
{\sl ${}^c$INFN, Sezione di Roma and Dipartimento di Fisica, \\
           Universit\`a di Roma ``La Sapienza'', P.le A. Moro 5, I-00185 Rome, Italy} \\ 
[25 pt]  }
   {\bf Abstract} \\
\end{center}\noindent
We analyse scalar flavour-changing neutral currents 
of down-type quarks in models with two Higgs doublets, 
coupled separately to up- and down-type quarks,
in the limit where the ratio of the two expectation 
values ($\tan \beta = v_u/v_d$) is large. We clarify the origin 
of this phenomenon, both in $\Delta F=1$ and 
$\Delta F=2$ processes, analysing  
differences and analogies between supersymmetric
and non-supersymmetric models. We confirm previous 
findings of a sizeable enhancement at large $\tan\beta$ 
of specific $\Delta F=1$ and $\Delta F=2$ 
amplitudes in the MSSM and, in these cases, we discuss 
how large-$\tan\beta$ corrections can be controlled
beyond lowest order. 
Finally, we emphasize the unique role of the rare 
processes $B_{s,d} \to \tau^+ \tau^-$ 
and $B_{s,d} \to \mu^+ \mu^-$
in probing this scenario.

\def\thefootnote{\arabic{footnote}}
\setcounter{footnote}{0}
\setcounter{page}{0}
\newpage

\section{Introduction}
Processes mediated by flavour-changing neutral-current (FCNC) 
amplitudes are extremely useful to deeply investigate  
the dynamics of quark-flavour mixing: 
the strong suppression of these transitions occurring within the 
standard model (SM), due to the absence of tree-level contributions 
and the hierarchy of the Cabibbo--Kobayashi--Maskawa (CKM) matrix, naturally
enhance their sensitivity to possible non-standard phenomena. 
This is even more true in the case of down-type FCNC amplitudes 
mediated by  scalar (and pseudoscalar) currents that, 
within the SM, are additionally suppressed by 
the smallness of down-type Yukawa couplings. 

The suppression of down-type FCNC scalar operators 
becomes less effective in models with an extended 
Higgs sector and, particularly, in the popular 
two Higgs doublets model (2HDM) of type II,   
where two $SU(2)_L$ scalar doublets are coupled 
separately to up- and down-type quarks (see e.g.~Ref.~\cite{Haber}).
In this case FCNC amplitudes are still absent at the 
tree level. However, contrary to the SM, it is possible to accommodate 
large down-type Yukawa couplings, provided the ratio $v_u/v_d=\tan\beta$,
where $v_u$ ($v_d$) denotes the vacuum expectation value
of the doublet coupled to up (down)-type quarks, is large.
Then, for instance, $\Delta F=1$ operators 
like $\bar b_R  s_L \bar \mu_R \mu_L$, suppressed by two small
Yukawa couplings, could receive a 
$\tan^2\beta$ enhancement with respect to the SM case.

Even though the $\tan^2\beta$  enhancement could be as large as 
$10^3$ in the simple 2HDM, scalar FCNC amplitudes
barely reach the level of vector-type  SM  amplitudes, 
and this happens only in very few cases. On the other hand, 
the 2HDM of type II is particularly interesting
being the Higgs sector of the Minimal Supersymmetric 
SM (MSSM) (see e.g. Refs.~\cite{Haber,Martin}). 
As first noted by Babu and Kolda \cite{Babu}, 
and later confirmed in \cite{Chanko,Bobeth},
the $\tan\beta$ enhancement of scalar 
FCNC amplitudes can be much more effective in the MSSM than in 
the non-supersymmetric case. In particular, 
the one-loop coefficient of the operator  $\bar b_R s_L \bar \mu_R \mu_L$
has been found to scale like $\tan^3\beta$, leading to 
possible large non-standard effects in $B_s \to \mu^+\mu^-$. 
More recently, it has also been shown that the  coefficient 
of the $\Delta F=2$ operator $\bar b_R s_L \bar b_L s_R$
receives, at the two-loop level,  
a contribution scaling as $\tan^4\beta$ that could have 
a relevant impact on $B_s$--${\bar B}_s$ mixing \cite{Buras}.

The purpose of this paper is to clarify the nature of this
$\tan\beta$ enhancements. We will analyse the large-$\tan\beta$ behaviour 
of all relevant down-type $\Delta F=1$ and $\Delta F=2$ amplitudes,
both in supersymmetric and non-supersymmetric cases.
As already pointed out in \cite{Babu}, the 
large-$\tan\beta$ enhancement of scalar FCNC amplitudes is intimately 
related to the appearance, at the one-loop level, of an effective 
coupling between $H_u$ and down-type quarks \cite{HRS}.
As we shall show, this is a necessary consequence of any 
2HDM and indeed it is realized in a very similar way in the 
supersymmetric and in the non-supersymmetric case.
The difference between the two scenarios arises only under a 
specific conspiracy of the soft-breaking terms, 
which could decouple
the supersymmetric ${\widetilde H}_u$--${\widetilde H}_d$ mixing
from the ordinary $H_u$--$H_d$ coupling,
necessarily suppressed in the large-$\tan\beta$ limit.
Analysing both scenarios, we shall clarify the origin 
of the different $\tan\beta$ factors, showing how to control
the large-$\tan\beta$ terms beyond lowest order.  

The paper is organized as follows. Section 2 is devoted 
to $\Delta F=1$ amplitudes; there we shall first analyse 
the generic structure of the effective $d_i-d_j-H^0$ vertex, 
then we shall discuss the large-$\tan\beta$ 
behaviour of $d_i \to d_j \ell^+ \ell^-$ amplitudes. 
In section 3 we analyse $B_{s,d}$--${\bar B}_{s,d}$ mixing, 
discussing the generic structure of both reducible 
and one-particle irreducible contributions and deriving 
phenomenological bounds from $\Delta M_{s,d}$.
Section 4 contains a phenomenological analysis 
of the rare decays $B_{s,d} \to \tau^+ \tau^-$ and 
$B_{s,d} \to \mu^+ \mu^-$. 
The results are summarized in the conclusions.

\section{$\Delta F=1$ scalar currents}
\label{sec:DF=1}
\subsection{The effective down-type Yukawa interaction}
In a 2HDM of type II (including the MSSM)
the tree-level Yukawa interaction is defined as
\beq
\cL_Y^0  =  \dbar_R Y_d   Q_L  H_d + \ubar_R Y_u   Q_L  H_u 
{\rm ~+~h.c.},
\eeq
where $Y_{u,d}$ are $3\times3$ matrices in flavour 
space and $H_{u,d}$ denote the two Higgs doublets.
This Lagrangian is invariant under a global 
$U(1)$ symmetry, which we shall call $U(1)_d$ and
under which $\dbar_R$ and $H_d$ have opposite 
charge and all the other fields are neutral. 
If this symmetry were exact, the coupling of 
$H_{u}$ to down-type quarks would be forbidden 
also at the quantum level. This symmetry, however, 
is naturally broken by terms appearing in the 
Higgs potential and, if $\tan\beta$ is large,
this has a substantial impact on the 
effective Yukawa interaction of down-type quarks. 
Under the assumption that $L_Y^0$ is the only source of 
flavour mixing and that the model-dependent 
Higgs self-couplings have the MSSM structure \cite{Haber,Martin}, 
the one-loop effective down-type Yukawa Lagrangian 
(in both the supersymmetric and non-supersymmetric cases) can 
be written as \cite{HRS,BRP,Babu}:
\beq
\cL_d^{\rm eff}  =  \dbar_R Y_d \left[ H_d 
+\left( \epsilon_0 + \epsilon_Y Y_u \da Y_u \right) H_u^* \right] Q_L 
{\rm ~+~h.c.}~,
\label{eq:l_eff}
\eeq 
where $\epsilon_{0,Y}$ denote appropriate
loop functions [$\epsilon_{0,Y} \sim \cO(1/16\pi^2)$]
proportional to the $U(1)_d$-breaking terms.

In order to diagonalize the mass terms generated by 
$\cL_d^{\rm eff}$, it is convenient to 
rotate the quark fields in the basis where $Y_d$ is diagonal 
[$(Y_d)_{ij}=  y^d_i \delta_{ij}  $]. In this basis we can write
\beq
\cL_{d-{\rm mass}}^{\rm eff}  
=  v_d   \dbar_R^i y^d_i  \left[ (1+ \epsilon_0 \tan \beta ) \delta_{ij}
+ \epsilon_Y \tan\beta\  V^{0\dagger}_{ik} (y^u_k)^2  V^{0}_{kj}   
\right] d^j_L {\rm ~+~h.c.}, 
\label{eq:l_mass}
\eeq
where $v_d^2 (1+\tan^2\beta) = (2\sqrt{2} G_F)^{-1} \approx (174~{\rm GeV})^2$ and 
$V^0$ denotes the tree-level CKM matrix, i.e.~the CKM matrix 
in the limit $\epsilon_Y=0$. 
Neglecting the small terms due to $y^{u}_{1,2}$,
defining $\lambda^t_{jk}=V^{0\dagger}_{j3} V^{0}_{3k}$ (for $j\not = k$)
and  $y_t = y^{u}_3$, we can rewrite Eq.~(\ref{eq:l_mass}) as
\beq
\cL_{d-{\rm mass}}^{\rm eff}  
=  v_d   \dbar_R^i {\bar y}^{d}_i  
\left[ \delta_{ij} + \frac{ \epsilon_Y y_t^2  \tan\beta }{ 
1 +   \tan \beta \left( \epsilon_0 + \epsilon_Y  y_t^2   |V^0_{i3}|^2 \right)}
 \lambda^t_{ij}
\right] d^j_L {\rm ~+~h.c.}~,
\label{eq:l_mass2}
\eeq
where
\beq
 {\bar y}^{d}_i =
y^{d}_i \left[ 1+ \tan \beta \left( \epsilon_0  + \epsilon_Y y_t^2 
|V^0_{i3}|^2 \right)\right]~ \approx~  
y^{d}_i \left[ 1+ \tan \beta \left( \epsilon_0  +  \epsilon_Y y_t^2 \delta_{i3}
 \right)\right]~. 
\label{eq:y_d_eff}
\eeq
Because of the hierarchy of the CKM matrix, $\lambda^t_{ij} \ll 1$ and 
we can diagonalize Eq.~(\ref{eq:l_mass2}) perturbatively in $\lambda^t_{ij}$.
The rotation that diagonalizes Eq.~(\ref{eq:l_mass2}) to the 
first order in $\lambda^t_{ij}$ is given by 
\beqa
d_L^j &\to& \left[  \delta_{jk} +   \epsilon_Y y_t^2  \tan\beta 
\frac{ {\bar y}^{d}_j {y}^{d}_j + {\bar y}^{d}_k {y}^{d}_k }{ 
 ({\bar y}^{d}_k)^2  -  ({\bar y}^{d}_j)^2 }  \lambda^t_{jk}
\right] d_L^k~, \label{eq:dLrot} \\
d_R^j &\to& \left[  \delta_{jk} +  \epsilon_Y y_t^2  \tan\beta 
\frac{ {\bar y}^{d}_j {y}^{d}_k + {\bar y}^{d}_k {y}^{d}_j }{ 
 ({\bar y}^{d}_k)^2  -  ({\bar y}^{d}_j)^2 } \lambda^t_{jk}
\right] d_R^k~. \label{eq:dRrot}
\eeqa
At this order the eigenvalues are not shifted, and thus  
the leading $\tan\beta$ corrections to the eigenvalues of the down-type 
Yukawa matrix are simply described by Eq.~(\ref{eq:y_d_eff}) \cite{HRS}.
On the other hand, the rotation (\ref{eq:dLrot}) modifies the 
structure of the CKM matrix, but only 
the $V_{3j}$ ($j\not=3$) elements receive $\cO(1)$ corrections 
(i.e. corrections not suppressed by the CKM hierarchy) \cite{BRP}:
\beq
\frac{  V_{3j} }{ V^0_{3j} } =
\frac{  V_{j3} }{ V^0_{j3} } =   \frac{ 1 + \epsilon_0  \tan\beta }{ 
1 +   \tan \beta (\epsilon_0 + \epsilon_Y  y_t^2)}~.
\eeq

As first noted by Babu and Kolda \cite{Babu}, if $\epsilon_Y \not=0$
the diagonalization of Eq.~(\ref{eq:l_mass2}) necessarily induces a 
FCNC coupling between quarks and neutral Higgs bosons. This can
easily be understood by looking at the neutral component
of Eq.~(\ref{eq:l_eff}): if $\epsilon_Y\not =0$ and $Y_u$ is
not aligned to $Y_d$, we have two independent flavour 
structures (not simultaneously diagonalizable) that are 
weighted differently for physical Higgses and mass terms.
Indeed performing explicitly the rotations (\ref{eq:dLrot}) and 
(\ref{eq:dRrot}) and neglecting 
the subleading $\cO( y^{d}_j/y^{d}_{k>j} )$ terms, 
we find 
\beq
\cL_{{\rm FCNC}~(k \not= j)}^{\rm eff}   
=  y^{d}_k {\bar \lambda}^t_{kj} 
 \frac{ \epsilon_Y y_t^2  \tan\beta }{ 1 + \epsilon_0 \tan \beta} 
\left[ \frac{1}{\tan \beta} H_u^{0*} - H_d^0   \right] \dbar_R^k d^j_L   {\rm ~+~h.c.}~,
\label{eq:L_FCNC}
\eeq
where, expressing the off-diagonal coupling in terms of the physical CKM matrix,
\beq
{\bar \lambda}^{t*}_{jk} = 
{\bar \lambda}^t_{kj} = \left\{ \ba{ll} 
 V^*_{k3}  V_{3j} \left[ \frac{ 1 + \tan \beta (\epsilon_0 + \epsilon_Y y_t^2) 
 }{ 1 + \epsilon_0 \tan \beta } \right]^2 \quad  &  (k\not=3,~j\not=3) \\
 V^*_{33}  V_{3j} & (k=3,~j\not=3)  \\
 V^*_{k3}  V_{33} \left[ \frac{ 1 +  \epsilon_0 \tan \beta  
 }{ 1 + \tan \beta(\epsilon_0 + \epsilon_Y y_t^2) } \right] \quad  & (k\not=3,~j=3) 
 \ea \right.
\label{eq:barlam_def}
\eeq
Equation (\ref{eq:L_FCNC}) generalizes to any $d^k \to d^j$ 
transition the result obtained by Babu and Kolda
for the $b \to s$ case. 

One of the most interesting aspects of Eq.~(\ref{eq:L_FCNC}) is the 
rapid growth with $\tan\beta$ of the $H^0_d \dbar_R^k d^j_L$ coupling.
Expressing $y^{d}_k$ in terms of down-type quark masses, 
the FCNC coupling grows almost quadratically in $\tan\beta$, 
contrary to the approximate linear growth of the diagonal 
 $H^0_d \dbar_R^k d^k_L$ term. This is the origin of the 
approximate $\tan^3 \beta$ and $\tan^4 \beta$ behaviour 
of the $\Delta F=1$ and $\Delta F=2$ amplitudes  
discussed in \cite{Babu} and \cite{Buras}, respectively.

As expected, the combination of neutral Higgs fields
that appears in Eq.~(\ref{eq:L_FCNC}) has a vanishing
vacuum expectation value and does not include the 
Goldstone component; it can therefore be expressed 
in terms of the three physical neutral Higgs states. 
Assuming the Higgs potential to be $CP$-invariant and 
employing the notation of \cite{Martin} we find
\beqa
\cL_{{\rm FCNC}~(k \not= j)}^{\rm eff}   
&=&  y^{d}_k {\bar \lambda}^t_{kj} 
 \frac{ \epsilon_Y y_t^2  }{ \sqrt{2} \cos \beta [1 + \epsilon_0 \tan \beta]  } \no \\
&& \times \left[ \cos(\alpha-\beta) h^0 + \sin(\alpha-\beta) H^0 - i A^0 \right] 
\dbar_R^k d^j_L   {\rm ~+~h.c.}
\label{eq:L_FCNC2}
\eeqa

\subsection{Explicit estimates of $\epsilon_Y$}
In the context of a non-supersymmetric 2HDM with
Higgs self-couplings fixed as in the MSSM case, the only 
source of   $U(1)_d$ breaking  is the bilinear operator 
$H_u H_d$. If the Higgs potential is $CP$-invariant,
the coupling of this operator, conventionally denoted by $\mu B$,
is related to the mass of the neutral pseudoscalar Higgs 
by the (tree-level) relation
\beq
 \mu B = M_A^2 \sin 2 \beta/2~.
\label{eq:Ma}
\eeq
Thus $U(1)_d$  breaking is unavoidable  
if we require a non-vanishing mass for $A^0$.  
Note, however, that in this case the $U(1)_d$ breaking 
is parametrically suppressed in the large-$\tan\beta$ limit.

\begin{figure}[t]
    \begin{center}
       \setlength{\unitlength}{1truecm}
       \begin{picture}(10.0, 5.5)
       \epsfxsize 10.  true cm
       \epsffile{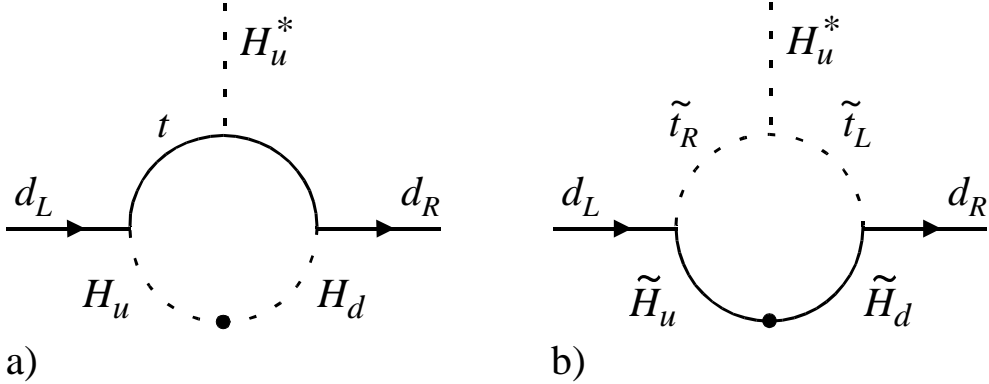}
       \end{picture}
    \end{center}
    \caption{Penguin diagrams generating a non-vanishing $\epsilon_Y$: 
             a) non-supersymmetric $H_u$--$H_d$ mixing; b) chargino mixing.}
    \protect\label{fig:1}
\end{figure}

The non-supersymmetric $H_u$--$H_d$ mixing induces a non-vanishing $\epsilon_Y$
via the mechanism shown in Fig.~\ref{fig:1}a. 
The computation of this effect at large $\tan\beta$
can be performed by setting $g=0$, i.e. switching off gauge interactions: 
keeping  $g\not=0$ would have complicated 
the calculation, owing to the gauge-dependent mixing of charged Higgs fields 
and $W$ bosons, without affecting the final result.
In this limit we can identify $H_u^+$ with the massless 
Goldstone boson and $H_d^+$ with the massive charged Higgs, and 
the only diagram to be computed is the one in Fig.~\ref{fig:1}a, which
leads to
\beq
\epsilon^{\rm 2HDM}_Y = \frac{1}{16 \pi^2} \frac{\mu B}{M_{H^+}^2}\left[ 
\log\left( m^2_t/M_{H^+}^2 \right) +\cO\left( m^2_t/M_{H^+}^2 \right)\right]~.
\label{eq:eps_2HDM}
\eeq

In the supersymmetric case an additional source of  $U(1)_d$ breaking  
is provided by the $\mu {\overline {\widetilde H}}_u {\widetilde H}_d$ term. 
The latter contribute to $\epsilon_Y$ via the diagram of Fig.~\ref{fig:1}b, 
leading to \cite{HRS}:
\beq
\epsilon^{\rm SUSY}_Y = \frac{1}{16 \pi^2} \frac{\mu A}{M_{\widetilde t_L}^2  } 
f\left( x_{\mu L} , x_{R L} \right)~,
\label{eq:eps_susy}
\eeq
where, as usual, $A$ denotes the coupling of the soft-breaking trilinear term 
(we assume both $\mu$ and $A$ to be real), 
 $x_{\mu L} = \mu^2/M_{{\widetilde t}_L}^2$ and 
 $x_{R L} =  M_{{\widetilde t}_R}^2/M_{{\widetilde t}_L}^2$.
The full expression of 
$f(x,y)$ can be found in the appendix and the normalization is such that $f(1,1)=1/2$. 
In the supersymmetric case one finds also a non-vanishing $\epsilon_0$, 
dominated by the contribution of gluino penguins \cite{HRS}.

As can be noted, the explicit expressions of $\epsilon^{\rm 2HDM}_Y$ and $\epsilon^{\rm SUSY}_Y$
are rather similar: $\epsilon_Y \sim 1/(16\pi^2)$ times an adimensional coupling 
parametrizing the $U(1)_d$ breaking. In the non-supersymmetric case 
the  $U(1)_d$ breaking is strongly constrained by  Eq.~(\ref{eq:Ma}), 
which forces  $\epsilon_Y$ to be suppressed as $1/\tan\beta$ in the 
large-$\tan\beta$ limit. 
Thus the potential $\tan^2\beta$ growth of the $H^0_d \dbar_R^i d^j_L$ coupling
cannot be realized in the simple 2HDM model. 
On the contrary, in the 
supersymmetric case we are allowed to consider a scenario where 
$\mu A/M_{\widetilde t_L}^2  = \cO(1)$ also at large $\tan\beta$.  
Note, however, that this scenario implies a sizeable hierarchy among 
soft-breaking terms: if $M_{{\widetilde t}_L} \gsim M_A$ we need 
$A/B \gsim \tan\beta$, while if $A/B \lsim 1$ we need 
$M^2_{{\widetilde t}_L}/M_A^2 \lsim \tan\beta$.

\subsection{$d_i \to d_j \ell^+\ell^-$ transitions}
We are now ready to discuss the effect of scalar-current amplitudes
in $d_i \to d_j \ell^+\ell^-$ transitions. 
The effective Hamiltonian describing these processes, 
including scalar-current operators, can be written as 
\beq
\cH_{\Delta F=1 }^{\rm eff} =  \cH_{\rm SM}^{\rm eff} +
C_S O_S  + C_P O_P + C'_S O'_S + C_P' O_P' {\rm ~+~h.c.}, 
\label{eq:Heff1}
\eeq
where $\cH_{\rm SM}^{\rm eff}$ denotes the SM 
basis of $\Delta F=1$ operators (see e.g.~\cite{Bobeth}) and
\beqa
&& O_S   = \dbar_R^i d^j_L \lbar \ell~, \qquad
O_P   =  \dbar_R^i d^j_L \lbar \gamma_5 \ell~, \no \\ && 
O_S'  = \dbar_L^i d^j_R \lbar \ell~, \qquad
O_P'  =  \dbar_L^i d^j_R \lbar \gamma_5 \ell~.
\label{eq:sc_ops}
\eeqa

One-loop contributions to  $b \to s,d$ transitions 
in the non-supersymmetric 2HDM have been discussed in 
detail in Refs.~\cite{Bobeth,Logan,Huang}. This analysis
requires, apparently, the evaluation of a large number of 
box and  penguin diagrams. 
However, the calculation can be strongly simplified by working 
in the limit $g=0$, which has proved to be the most convenient 
framework to discuss leading $\tan\beta$ contributions to
scalar FCNC amplitudes. 
In this limit it is easy to realize that box diagrams can be neglected, 
as long as we ignore terms suppressed by two lepton Yukawa couplings.

\begin{figure}[t]
    \begin{center}
       \setlength{\unitlength}{1truecm}
       \begin{picture}(10.0, 5.5)
       \epsfxsize 10.  true cm
       \epsffile{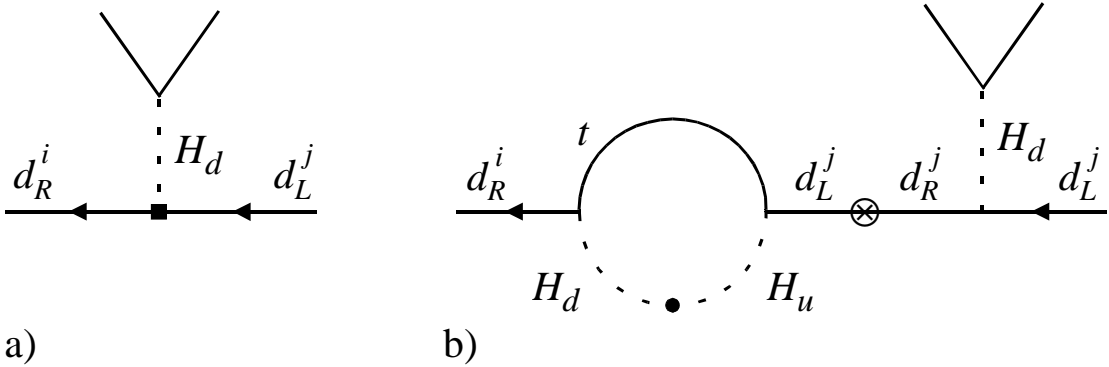}
       \end{picture}
    \end{center}
    \caption{Feynman diagrams relevant to the $d^i_R \to d_L^j \ell_L \ell_R$ amplitude: 
             a) leading contribution computed by means of the effective Yukawa interaction;
             b) one-loop contribution to the mass mixing in the 2HDM.}
    \protect\label{fig:2}
\end{figure}

Penguin (and self-energy) diagrams can be evaluated by means of  the 
effective Yukawa interaction discussed before, as shown in Fig.~\ref{fig:2}a.
The limitation of the latter is that only the leading 
dimen\-sion-four operator $H^{*}_u \dbar_R Q_L$ has been considered. 
Higher-dimensional operators with more powers of $H_u$, which
contribute to the mass diagonalization when $H_u$ acquires a v.e.v.,
have been neglected. The coefficients of these 
higher-dimensional operators are necessarily 
suppressed by the 
inverse power of the heavy scale of the theory, namely the charged Higgs mass
in the non-supersymmetric 2HDM.
Therefore, using $\cL_{\rm FCNC}^{\rm eff}$, 
we can control only the leading term in an expansion in 
powers of $v_u/M_{H^+}$. This leads to the following initial conditions 
for the Wilson coefficients in (\ref{eq:Heff1}):
\beqa
 C_S = \ \  C_P &=&  \frac{ y^d_i y_\ell y_t^2 {\bar \lambda}^t_{ij} }{32 \pi^2 } \times \frac{1}{M_{H^+}^2} 
 \left[ \log\left( \frac{M_{H^+}^2}{m^2_t} \right) +\cO\left( \frac{m^2_t}{M_{H^+}^2} \right)\right]~, \no\\
 C'_S = - C'_P &=&  \frac{ y^d_j y_\ell y_t^2 {\bar \lambda}^t_{ji} }{32 \pi^2}  \times \frac{1}{M_{H^+}^2} 
 \left[ \log\left( \frac{M_{H^+}^2}{m^2_t} \right) +\cO\left( \frac{m^2_t}{M_{H^+}^2} \right)\right]~,
\label{eq:CSCP}
\eeqa
where subleading $\cO(1/\tan\beta)$ terms have been neglected.
These results are compatible with those in Refs.~\cite{Bobeth,Logan,Huang}
and, even though $\cO({m^2_t}/{M_{H^+}^2})$ corrections are missing, 
Eqs.~(\ref{eq:CSCP}) have the advantage of incorporating higher-order $\tan\beta$ 
terms, hidden in $y^d_i$ and ${\bar \lambda}^t_{ij}$ 
[see Eqs.~(\ref{eq:y_d_eff}) and (\ref{eq:barlam_def})]. 
As long as we neglect terms suppressed by additional down-type 
Yukawa couplings and/or additional off-diagonal CKM factors, 
these are the only source of $\tan\beta$-enhanced corrections.

It is worthwhile to stress that the subleading $\cO({m^2_t}/{M_{H^+}^2})$
terms in Eqs.~(\ref{eq:CSCP}) can also be computed 
in a rather simple way. To this purpose we only need to perform a
complete diagonalization of the off-diagonal mass term generated at 
one loop, which is equivalent to computing the diagram in Fig.~\ref{fig:2}b
(in the gaugeless limit).
As a result, we find that the term within square brackets in 
 Eqs.~(\ref{eq:CSCP}) should be replaced by
\beq
\frac{M_{H^+}^2 }{M_{H^+}^2 -m^2_t} \log\left( \frac{M_{H^+}^2}{m^2_t} \right)~,
\label{eq:CSCP_bis}
\eeq
in agreement with  the results of Refs.~\cite{Bobeth,Logan,Huang}.

Full one-loop calculations in the supersymmetric case 
have been performed in Ref.~\cite{Chanko,Bobeth}, 
complementing the result of Ref.~\cite{Babu} obtained by 
means of the effective Yukawa couplings. As expected, 
the latter provides an excellent approximation to the 
full result in the limit of heavy squark masses. 
The supersymmetric contribution to $C_S=C_P$ obtained by means 
of  $\cL_{\rm FCNC}^{\rm eff}$ can be written as 
\beqa
 C_S &=& -
\frac{ y^d_i y_\ell y_t^2 {\bar \lambda}^t_{ij} }{32 \pi^2 } \times 
\frac{A   }{B  M_{\widetilde t_L}^2  [1+\epsilon_0 \tan\beta] } f\left( x_{\mu L} , x_{R L} \right) ~, 
\label{eq:susy1}\\
 &=& - \frac{G_F^2}{ 4 \pi^2 } \times \frac{ m_{d_i}  m_\ell m_t^2 {\bar \lambda}^t_{ij}  \tan^2\beta  }{
 1+ \tan \beta ( \epsilon_0  +  \epsilon_Y y_t^2 \delta_{i3} )  } 
\times \frac{\mu A \tan\beta }{ M_{\widetilde t_L}^2  M_A^2  [1+\epsilon_0 \tan\beta] } 
f\left( x_{\mu L} , x_{R L} \right)~.
\label{eq:susy2}
\eeqa
The comparison between (\ref{eq:susy1}) and (\ref{eq:susy2}) 
illustrates the origin of the various $\tan\beta$ factors.
Some comments are in order:

\begin{enumerate}

\item{} The third $\tan\beta$ factor 
in the numerator is not directly related to the 
down-type Yukawa couplings. 
As anticipated, with a different choice of parameters it 
can be eliminated in favour of the ratio $A/B$.

\item{} Higher-order corrections enhanced by $\tan\beta$, 
shown explicitly in Eq.~(\ref{eq:susy2}), 
cannot be neglected (similarly to the $b\to s \gamma$ 
case \cite{bsgamma}). For $\tan\beta \gsim 30$ these 
higher-order terms are numerically 
more important than the remaining one-loop corrections
not described by  $\cL_{\rm FCNC}^{\rm eff}$.
By means of Eq.~(\ref{eq:susy2}) we take into 
account all terms of the type 
$(G_F m_b m_\ell) \times ({\bar \alpha}/\pi)^{n+1} (\tan\beta)^{n+3}$,
where $n \geq 0$ and ${\bar \alpha}$ denotes either  $\alpha_S$
or $y_t^2/(4\pi)$.

\item{} The $\tan\beta$-enhanced corrections 
to the $\bar t_R b_L H^+$ vertex, which play an important 
role in $b\to s \gamma$ although formally 
subleading \cite{bsgamma}, are not 
relevant here owing to the strong overall suppression of the
charged-Higgs contribution with respect to the chargino one.

\end{enumerate}

\medskip
We conclude this section with a generic comparison 
between scalar FCNC amplitudes at large $\tan\beta$, 
both in the 2HDM and in the MSSM, and ordinary SM 
vector-type amplitudes.
On general grounds, the former 
are suppressed with respect to the latter by a factor 
\beq
\frac{M^2_W}{M_A^2} \times  \frac{ m_{d_i} m_\ell}{ m_t^2 } \times \tan^n\beta~,
\label{eq:fact}
\eeq
where $n \approx 2$ in the non-supersymmetric case and  $n \lsim 3$ in the MSSM. 
An exception to this rule is provided by $P\to \ell^+\ell^-$ decays, 
discussed in detail in section~4; here the 
helicity suppression of the SM matrix element leads to 
replacing the factor $ m_{d_i}m_\ell$ in (\ref{eq:fact}) by $m^2_{d_i}$.
Taking into account the approximate numerical relations 
\beq
\frac{ m_s }{ m_t }~,~
\frac{ m_\mu }{ m_t }~\sim~\cO\left( \frac{1}{\tan^2\beta} \right) \qquad {\rm and} \qquad
\frac{ m_b }{ m_t }~,~  
\frac{ m_\tau }{ m_t }~\sim~\cO\left( \frac{1}{\tan\beta} \right)
\label{eq:powerc}
\eeq  
that hold for $30 \lsim \tan\beta \lsim 50$,
the general picture can be summarized as follows.
\begin{description}
\item{$\underline{b \to (s,d) \tau^+\tau^-}$:} non-standard scalar  contributions 
could have a sizeable impact already at the level of inclusive transitions. 
The magnitude of the scalar amplitude could be substantially larger than the SM one 
in the  MSSM, whereas it is typically smaller  than the SM one in the non-supersymmetric case.
\item{$\underline{b \to (s,d) \mu^+\mu^-}$:} 
the impact of scalar currents 
is almost negligible in inclusive transitions 
(as explicitly shown in Ref.~\cite{Bobeth}),
but large effects are still
possible in the exclusive dilepton decays, especially within the MSSM.
Indeed  the most stringent constraint on 
$\Delta F=1$ scalar currents --under the hypothesis of minimal flavour violation--
is at present derived from the experimental bound on $B(B_s \to \mu^+\mu^-)$
(see section~4).
\item{$\underline{s \to d \mu^+\mu^-}$:}  
scalar current contributions could reach at most a
few percent of the SM short-distance amplitude in $K_L \to \mu^+\mu^-$. 
Given the theoretical uncertainties affecting long-distance
contributions to this mode \cite{KL}, these non-standard effects are not detectable.
We stress, however, that this conclusion holds only under the 
hypothesis of minimal flavour violation. If the CKM matrix is not 
the only source of flavour mixing, the $\epsilon_{Y,0}$ parameters 
are not flavour-diagonal and the  $H^0_d \dbar_R^i d^j_L$ coupling
is not necessarily proportional to $\lambda^t_{ij}$. Within this more general
framework, it is then possible to overcome the strong $\lambda^t_{12}$ 
suppression of $s \to d \mu^+\mu^-$  amplitudes and to generate $\cO(1)$
effects also in $K$ decays. 
A detailed analysis of this scenario is beyond the purpose of this paper,
where we shall restrict our attention to the hypothesis of minimal flavour violation
and, henceforth, to $B$ physics.
\end{description}

\section{$B_{d,s}$--${\bar B}_{d,s}$ mixing}
\label{sec:BB}
\subsection{Generalities}
Within new-physics scenarios where flavour mixing is 
governed only by the CKM matrix, the 
effective Hamiltonian relevant to $B_{d,s}$--${\bar B}_{d,s}$ 
mixing can be written as 
\beq
H^{\Delta B = 2}_{\rm eff} =  \frac{G^2_F}{16 \pi^2} M^2_W \left[ V_{tb}^{*}V_{t(d,s)} \right]^2  
        \sum_{i}C_{i}(\mu) Q_{i}  {\rm ~+~h.c.}, 
\eeq
where the full basis of dimension-six operators can be found, for example 
in Ref.~\cite{DB2RGE}. In addition to the SM operator
\beq
Q^{VLL} = {\bar b}_L \gamma_{\mu}  q_L 
              {\bar b}_L \gamma^{\mu}  q_L ~,
\label{eq:ops1}
\eeq
those relevant at large $\tan \beta$ are given by
\bea
Q^{SLL} &=& {\bar b}_R  q_L 
              {\bar b}_R  q_L~,  \nnu \\
Q^{SLR} &=&  {\bar b}_R  q_L  
              {\bar b}_L  q_R~,   \nnu \\
Q^{VRR} &=& {\bar b}_R \gamma_{\mu} q_R 
              {\bar b}_R \gamma^{\mu} q_R~,
\label{eq:ops2}
\eea
where $q=d,s$. 

\begin{figure}[t]
    \begin{center}
       \setlength{\unitlength}{1truecm}
       \begin{picture}(10.0, 5.0)
       \epsfxsize 10.  true cm
       \epsffile{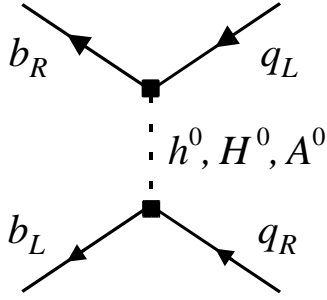}
       \end{picture}
    \end{center}
    \caption{Double-penguin diagram contributing to $C^{SLR}$; the black boxes 
     denote the effective FCNC Yukawa interaction of Eq.~(\ref{eq:L_FCNC2}).}
    \protect\label{fig:dp}
\end{figure}

All the Wilson coefficients of the four operators in Eqs.~(\ref{eq:ops1})
and (\ref{eq:ops2}) 
receives one-particle irreducible one-loop contributions from box diagrams. 
However, as pointed out in Ref.~\cite{Buras}, in the case of $Q^{SLR}$ 
a sizeable contribution at large $\tan\beta$ 
is also provided by the reducible two-loop diagram in Fig.~\ref{fig:dp}.

The Wilson coefficients of the three operators in Eqs.~(\ref{eq:ops2}) 
are all suppressed by small down-type masses, corresponding to the right-handed fields,
but are possibly enhanced by appropriate $\tan\beta$ factors,
as we shall discuss in the following. The smallness of $m_d$ 
implies that only $Q^{SLL}$ plays a significant role in ${\Delta M}_{d}$.
On the other hand, both $Q^{SLL}$ and $Q^{SLR}$
are potentially relevant to  ${\Delta M}_{s}$. 
Despite a $\tan^4\beta$
enhancement, the contribution of $Q^{VRR}$ is practically
negligible for both  ${\Delta M}_{d}$ and  ${\Delta M}_{s}$.

Following the notation of Ref.~\cite{Buras},
the $B_{d,s}$--${\bar B}_{d,s}$ mass difference 
can be written as
\bea
{\Delta M }_{d,s}=\frac{G^2_F M^2_W }{6 \pi^2} \eta_B m_{B_{d,s}} 
        ({\hat B}_{B_{d,s}} F^2_{B_{d,s}} ) F^{d,s}_{tt}
         |V_{t(d,s)}|^2\; ,  
\eea
where $({\hat B}_{B_{d,s}} F^2_{B_{d,s}})$ parametrize the 
hadronic matrix elements, to be determined through non-perturbative methods, 
and $\eta_B$ is the QCD renormalization-group factor of the SM operator, 
given by $\eta_B= 0.55$  at next-to-leading order (NLO).   
The real functions $F^{d,s}_{tt}$, encoding short-distance contributions, 
are conveniently decomposed as
\bea
F^{d,s}_{tt}=S_0(x_{tW})[1+f_{d,s}]~,
\eea
where $x_{tW}= m^2_t/M^2_W$, $S_0(x_{tW})$ accounts for the SM box
(see appendix)
and all non-standard effects are included in $f_{d,s}$. 
Expressing $f_{d,s}$ in terms of the (non-standard) 
contributions to the Wilson coefficients 
of the four-fermion operators in Eqs.~(\ref{eq:ops1}) and (\ref{eq:ops2}),
we can write 
\beq
f_{d,s} = \frac{1}{4 S_0(x_{tW})}\left[ C^{VLL} (\mu_t)+ 
         C^{VRR} (\mu_t) 
        + 4 {\bar P}^{SLR} C^{SLR} (\mu_t) 
        + 4 {\bar P}^{SLL} C^{SLL}(\mu_t)\right]~,
\label{eq:fds}
\eeq
where $\mu_t =\cO(m_t)$ and the explicit expressions 
of the ${\bar P}^{i}$, taking 
into account renor\-ma\-li\-za\-tion-group QCD corrections \cite{DB2RGE} 
and matrix elements of the scalar operators 
--normalized to the SM one-- can be found in 
Ref.~\cite{Buras}.\footnote{In principle the QCD correction factor
of non-standard vector operators is not exactly  $\eta_B$;
however, this difference can be ignored at the level of accuracy 
we are interested in. Similarly, we will neglect all finite QCD 
corrections to the initial conditions of the non-standard 
Wilson coefficients and corrections due to the running  
between $\mu_t$ and the new-physics scale.} 
Assuming $B$-parameters of the scalar operators 
equal to one, setting  $m_b(\mu_b) = 4.2$ GeV, $\alpha_s(M_Z)=0.118$
and ${\hat B}_{B_{d,s}}=1.3 \pm 0.2$ we find
\beq
4 {\bar P}^{SLR}=3.5\pm0.5~,  \qquad  4 {\bar P}^{SLL}=-2.1\pm0.3~.
\eeq
As can be noted, QCD effects enhance the contribution of 
scalar operators, especially $Q^{SLR}$, 
thus even small new physics contributions to their Wilson 
coefficients may be relevant to the phenomenological analysis.

\subsection{$\Delta B=2$ box diagrams}
Computing explicitly non-supersymmetric box diagrams with the 
exchange of $W^\pm$ and $H^\pm$ or two $H^\pm$ (Goldstone bosons 
included), in the large-$\tan\beta$ limit, we obtain the 
following initial conditions for the Wilson coefficients 
\bea
C_{H}^{VLL}  &=& \frac{2}{\tan^2{\beta}} \;[L_2(x_{tW},x_{tW},x_{HW}) -
          4 L_1(x_{tW},x_{tW},x_{HW})] \; ,\nnu \\ 
C_{H}^{VRR}  &=& \frac{m^2_b  m^2_{d,s} \tan^4\beta }{M^2_W M^2_{H^\pm}} 
          \; [ L_3(x_{tH},x_{tH},1)-2 L_3(x_{tH},0,1)+ L_3(0,0,1)] 
          \; ,\nnu \\
C_{H}^{SLL}  &=& \frac{ 4m^2_b}{ M_W^2} \; [L_1(x_{tW},x_{tW},1) + 
          L_1(x_{tH},x_{tH},1) - 2 L_1(x_{tH},x_{tH},x_{WH}) ] \;,\nnu  \\
C_{H}^{SLR}  &=& \frac{ 8 m_b  m_{d,s} \tan^2{\beta}    }{ M^2_{H^\pm} }
          \; [ - L_3(x_{tH},x_{tH},x_{WH})+ 2 L_3(x_{tH},0,x_{WH}) \nnu\\
           & & \qquad\qquad\qquad\quad
          - L_3(0,0,x_{WH}) + x_{HW} L_1(x_{tH},x_{tH},x_{WH})] 
          \; ,  
\eea
where, as usual, $x_{ab}= M^2_a/M^2_{b}$ and the loop 
functions $L_i(x,y,z)$ are defined in the appendix. 
These results are in agreement with those recently 
reported in Ref.~\cite{Buras}.

The charged-Higgs contribution to 
$C^{VLL}_H$ is directly suppressed 
by two inverse powers of $\tan \beta$ and turns out to be 
completely negligible. The strong $\tan\beta$ enhancement of 
$C_{H}^{VRR}$ is more than compensated by the down-type quark-mass terms:
employing the power counting in Eq.~(\ref{eq:powerc}), $C_{H}^{VRR}$
is effectively of $\cO(1/\tan^2\beta)$ at most, and indeed it is numerically 
irrelevant. Similarly, also $C_{H}^{SLL}$ is negligible being 
effectively of $\cO(1/\tan^2\beta)$.
The potentially largest contribution --still suppressed 
with  respect to the SM one-- is provided by 
$C^{SLR}_H$, which in the $B_{s}$--${\bar B}_{s}$ 
case is effectively suppressed 
by only one inverse power of $\tan\beta$. 
For $M_H \approx m_t$, $C^{SLR}_H$ 
can induce at most a $10\%$ correction to  $F^{s}_{tt}$:
\bea 
\delta f_s^{\rm 2HDM-box} &=&
\frac{{\bar P}^{SLR}_2 C^{SLR}_H}{ S_0(x_{tW})}
        = \frac{m_b m_{s} \tan^2 \beta}{M^2_W} \left[- 1.5-(x_{tH}-1)+
        {\cal O}(x_{tH}-1)^2 \right]  \no \\
        &\approx & \left[ -0.10-0.07(x_{tH}-1)+
        {\cal O}(x_{tH}-1)^2\right] \left(  \frac{\tan\beta}{50} \right)^2 
\eea
[$m_b(\mu_t)\approx 3$ GeV, $m_s(\mu_t)\approx 60$ MeV].
We can therefore conclude that non-supersymmetric 2HDM 
box contributions at large $\tan\beta$ 
do not induce appreciable effects in 
$B^0$--${\bar B}^0$ mixing. 

\begin{figure}[t]
    \begin{center}
       \setlength{\unitlength}{1truecm}
       \begin{picture}(10.0, 4.0)
       \epsfxsize 10.  true cm
       \epsffile{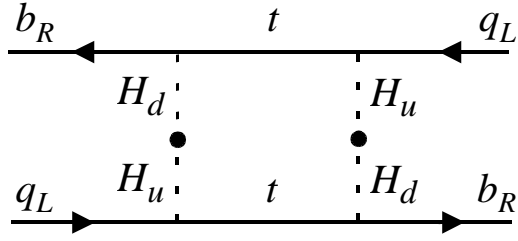}
       \end{picture}
    \end{center}
    \caption{Leading box-diagram contribution to $C^{SLL}$ in the non-supersymmetric
             2HDM.}
    \protect\label{fig:box}
\end{figure}

An important point to note is the following:
whereas each down-type mass term is compensated 
by a corresponding $\tan\beta$ factor, as naively 
expected, in $C_{H}^{SLR}$ and $C_{H}^{VRR}$, 
this does not occur in $C_{H}^{VLL}$ and $C_{H}^{SLL}$. 
In the case of  $C_{H}^{SLL}$, the reason 
of the additional $1/\tan^2\beta$ suppression 
is its vanishing in the absence 
of $U(1)_d$ breaking. Similarly to the case of 
$\Delta B=1$ amplitudes, this fact can easily 
be understood in the gaugeless limit. As shown 
in Fig.~\ref{fig:box}: the two $H_u$--$H_d$ mixing 
terms compensate the two $\tan\beta$ 
factors of the Yukawa couplings and, as a result, 
$C_{H}^{SLL}$ has no explicit $\tan\beta$ dependence. 
As shown in section~\ref{sec:DF=1}, 
$U(1)_d$ breaking is not necessarily suppressed 
at large $\tan\beta$ in the supersymmetric case,
when the $H_u$--$H_d$ mixing is replaced by the 
${\widetilde H}_u$--${\widetilde H}_d$ one.
We can therefore expect a $\tan^2\beta$-enhanced 
chargino-squark contribution to $C^{SLL}$.

The full one-loop contributions to  $C^{VLL}$ and $C^{SLL}$
generated by  chargino-squark diagrams in the MSSM can be written as 
\bea       
C_{\chi}^{VLL} &=& 4[f(\tilde{u},\tilde{u}) -2 f(\tilde{u},\tilde{t}) +
          f(\tilde{t},\tilde{t})]\;   ,    \label{eq:fff}    \\ 
C_{\chi}^{SLL} &=& 4[g(\tilde{t},\tilde{t}) -2 g(\tilde{u},\tilde{t}) +
          g(\tilde{t},\tilde{t})]\; ,  
\eea
where 
\bea 
f(\tilde{u},\tilde{t}) &=&
          \sum_{i,j,h,k=1,2} x_{W \chi_j} Y_{i \tilde{u}_h }Y_{i\tilde{t}_k}
          Y_{j\tilde{t}_k} Y_{j\tilde{u}_h } 
          L_3(x_{\tilde{t}_k \chi_j},x_{\tilde{u}_h \chi_j},x_{\chi_i\chi_j})~, 
          \\
g(\tilde{u},\tilde{t}) &=& 4 \sum_{i,j,h,k=1,2} x_{W \chi_j} 
          \frac{M_{\chi_i}}{M_{\chi_j}}
          Y_{i\tilde{u}_h} Z^b_{i\tilde{t}_k}
          Y_{j\tilde{t}_k}  Z^b_{j\tilde{u}_h }
          L_4(x_{\tilde{t}_k \chi_j},x_{\tilde{u}_h \chi_j},x_{\chi_i\chi_j}) ~,
\label{eq:ggg}
\eea
and $Y_{j\tilde{u}_h }$ and $Z^q_{j\tilde{u}_h }$ are defined as in 
Ref.~\cite{GabrielliGiudice}.\footnote{~The superscript in
$Z^q_{j\tilde{u}_h }$ refers to the down-type Yukawa coupling 
involved.}

The chargino-squark contribution to the Wilson coefficient of the 
standard vector operator ($C_{\chi}^{VLL}$), which includes the 
supersymmetrization of the ordinary $W$-box diagram,
has no special features at large $\tan\beta$: it is not 
suppressed or enhanced. For this reason, and also because 
$C_{\chi}^{VLL}$ has been extensively discussed in the literature 
(see, in particular, Refs.~\cite{GabrielliGiudice,Bertolini}),
we will not discuss it further. 

Contrary to $C_{\chi}^{VLL}$, the chargino-squark contribution to the
scalar operator $Q^{SLL}$ has not been discussed before and, as anticipated, 
it has an interesting behaviour at large $\tan\beta$.  In order to provide a
compact analytical expression and a numerical evaluation of 
$C_{\chi}^{SLL}$, we employ the following simplifying assumptions:
\begin{itemize}
\item we neglect up and charm Yukawa couplings and any non-degeneracy 
among the first two generations of squarks:
$M_{{\tilde q}_i}={\tilde M}$ for $q=u,c$ and $i=1,2$ [as already 
 implicitly assumed in Eqs.~(\ref{eq:fff})--(\ref{eq:ggg})]; 
\item we neglect off-diagonal terms in the chargino mass matrix
(or we assume $M_W \ll {\rm~max}\{ M_2,\mu \} )$, so that $M_{\chi_1}\approx 
M_2$ and $M_{\chi_2}\approx \mu$;
\item we neglect the left--right mixing of the squarks, but in the stop sector;
there we introduce a mixing angle $(-\pi/2<\theta_{\tilde t}
<\pi/2)$ satisfying the relation
\bea
\sin{2 \theta_{\tilde t}}=\frac{2 m_t(A-\mu \cot \beta)}{M^2_{{\tilde t}_1}
- M^2_{{\tilde t}_2}} \approx \frac{2m_t A }{M^2_{{\tilde t}_1}- M^2_{{\tilde t}_2}}
~,
\eea
where the ${\tilde t}_{1}$ mass eigenstate  ($M_{{\tilde t}_1}<M_{{\tilde t}_2}$)
can be identified with the right-handed stop for small  $\theta_{\tilde t}$~.
\end{itemize}
In the limit where $M_{{\tilde t}_2}$ is degenerate with the 
first two generations of squarks, 
while  $M_{{\tilde t}_1}$ is kept lighter, we obtain the following rather simple
expression:
\bea   
g(\tilde{t},\tilde{t}) &=& \frac{y_t^2 y_b^2  \sin^2 {2 \theta_t}}{32  G_F^2 M_W^2 \mu^2}
         ~G(x_{\tilde{t_1} \mu} ,x_{\tilde{t_2} \mu})\qquad\quad 
\left[ g(\tilde{u},\tilde{t})=g(\tilde{u},\tilde{u})=0 \right]~,
\label{eq:gs1}
\eea
where
\bea
G(x,y) &=& L_4(x,x,1)- 2L_4(x,y,1)+ L_4(y,y,1) \no \\
       &=& \frac{ G_2(y) }{y^2 } (x-y)^2  + {\cal O}(x-y)^3 ~,
\eea
and $G_2(y)$ can be found in the appendix. 
The expansion of $G(x,y)$ around $x\approx y$ lets us 
further simplify Eq.~(\ref{eq:gs1}), eliminating $\sin 2 \theta_{\tilde t}$
in favour of $A$. We then obtain
\bea   
g(\tilde{t},\tilde{t}) &=& \frac{ m_t^4 m^2_b \tan^2 \beta}{ M_W^2 \mu^4          
       \left[  1+ \tan \beta ( \epsilon_0  +  \epsilon_Y y_t^2 )\right]^2 } 
         \left(\frac{\mu A}{M_{\widetilde t_L}^2  } 
          \right)^2 G_2( x_{L \mu})\left[ 1+ \cO(1-x_{LR}) \right]~,
\label{eq:gs2}
\eea
where the Yukawa couplings have been expressed in terms 
of quark masses. Note the analogy of Eq.~(\ref{eq:gs2})
to Eq.~(\ref{eq:eps_susy}): in both cases the overall 
effect is ruled by the ratio $\mu A /M_{\widetilde t_L}^2$, 
which appears squared in the $\Delta B=2$ amplitude.
However, since the effective FCNC interaction of 
Eq.~(\ref{eq:L_FCNC}) 
does not enter the box amplitude, the $\tan\beta$ enhancement 
of the latter simply scale with the power of down-type Yukawa couplings.
As we shall discuss later on, 
for large $\tan\beta$, small $\mu$ ($\mu \lsim 200$~GeV)
and $\mu A/M_{\widetilde t_L}^2  = \cO(1)$,
this supersymmetric scalar contribution has a small but 
non-negligible impact on ${\Delta M}_{d}$ and  ${\Delta M}_{s}$.

\subsection{Double-penguin diagrams}
Double-penguin diagrams are formally a higher-order  
(two-loop) effect; however, the large-$\tan\beta$
growth of the effective Yukawa coupling make them 
numerically competing with one-loop contributions 
in the $B_s$--$\bar B_s$ case \cite{Buras}. 

The structure of Eq.~(\ref{eq:L_FCNC2}) leads, in general, to 
a vanishing result when the Higgs fields are contracted 
to generate effective operators of the type 
${\bar b}_R q_L {\bar b}_R q_L$ or 
${\bar b}_L q_R {\bar b}_L q_R$ ($q=s,d$). 
On the other hand, 
a non-vanishing result is obtained for effective operators
of the type ${\bar b}_R q_L {\bar b}_L q_R$, or when the 
combination of  Higgs fields explicitly shown in 
Eq.~(\ref{eq:L_FCNC2}) is contracted with its Hermitian conjugate. 
As a result, only $C^{SLR}$ is affected by 
double-penguin contributions. 

\begin{figure}[t]
    \begin{center}
       \setlength{\unitlength}{1truecm}
       \begin{picture}(10.0, 9.0)
       \epsfxsize 10.  true cm
       \epsffile{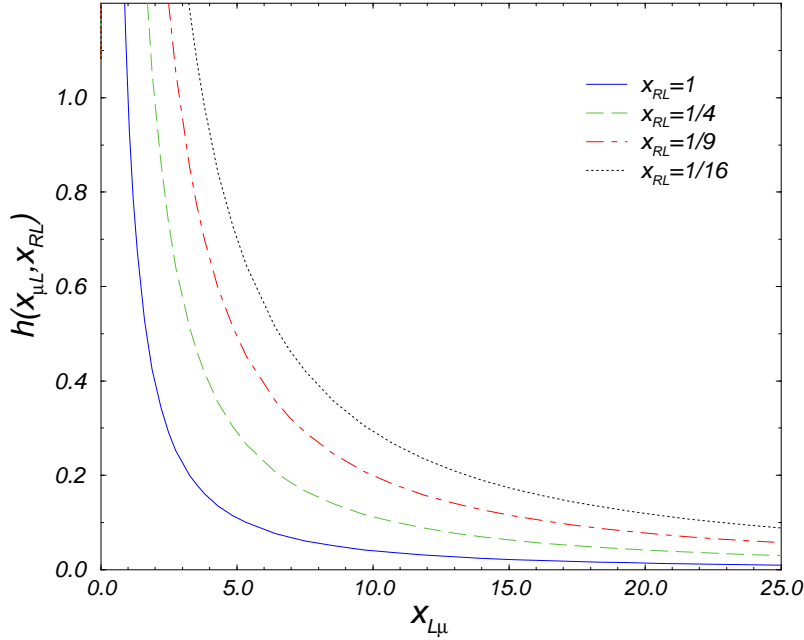}
       \end{picture}
    \end{center}
    \caption{$h(x_{\mu L},x_{R L})$ as a 
             function of $x_{L\mu}$ $(=1/x_{\mu L})$ for different values
             of $x_{R L}$.}
    \protect\label{fig:h}
\end{figure}

In general, the double-penguin contribution
to $C^{SLR}$ at large $\tan\beta$ can be written as 
\bea
C^{SLR}= - \frac{8 \pi^2 y_by_{s,d} }{G_F^2 M_W^2  } 
 \left[ \frac{ 1 +  \epsilon_0 \tan \beta  
 }{ 1 + \tan \beta(\epsilon_0 + \epsilon_Y y_t^2) } \right]
 \left( \frac{ \epsilon_Y y^2_t \tan\beta }{1+\epsilon_0 \tan \beta } 
           \right)^2 {\cal F}_+~,
\eea
where \cite{Buras}
\bea
{\cal F}_+ &=& \frac{\cos^2(\alpha - \beta)}{M^2_{h^0}} +
           \frac{\sin^2(\alpha - \beta)}{M^2_{H^0}} +
           \frac{1}{M^2_{A^0}}~\stackrel{M_A^2 \gg M_W^2 }{ 
           \longrightarrow }~\frac{2}{M^2_A}
           \; .
\eea
Within the non-supersymmetric case, the $1/\tan\beta$ 
suppression of $\epsilon^{\rm 2HDM}_Y$ leads to
negligible effects for both  ${\Delta M}_{d}$ and  
${\Delta M}_{s}$. On the contrary, the
supersymmetric contribution to ${\Delta M}_{s}$
can compete with the SM one. Expressing Yukawa 
couplings in terms of quark masses, we find
\bea
C_{\chi}^{SLR}  &=& - 
          \frac{G_F m_b m_s m^4_t}{ \sqrt{2} \pi^2 M^2_W }
          \frac{\tan^4 \beta  {\cal F}_+ 
          f^2(x_{\mu L},x_{R L}) }{ [1+\epsilon_0 \tan \beta]^2
            [  1+ \tan \beta ( \epsilon_0  +  \epsilon_Y y_t^2 ) ]^2 }
          \left(\frac{\mu A}{M_{\widetilde t_L}^2  } \right)^2 \; .
\label{eq:CSLRd}
\eea
The $\tan^4\beta$ enhancement of this coefficient is partially compensated by the 
$m_s$ suppression and the additional $1/16\pi^2$ (due to the two-loop order). 
As a result, it is of the same order as the chargino-stop box 
contribution to $C_{\chi}^{SLL}$. Neglecting higher-order 
$\tan\beta$ terms one finds 
\bea  
\frac{C^{SLL}_\chi}{C^{SLR}_\chi} & \approx & \frac{4\pi^2}{5\sqrt{2}} \times 
\frac{ m_b  M_A^2  h(x_{\mu L},x_{R L}) }{m_s  \mu^4 G_F \tan^2\beta }~ \no\\
&\approx& 
\left(\frac{M_A}{\mu}\right)^2 \left( \frac{\rm 100~GeV}{\mu} \right)^2
\left( \frac{50}{\tan\beta} \right)^2  h(x_{\mu L},x_{R L})~,
\label{eq:LLvLR}
\eea
where the function $h(x_{\mu L},x_{R L})$, normalized to $h(1,1)=1$,
is plotted in Fig.~\ref{fig:h}.

\subsection{Numerical bounds from $\Delta M_{d,s}$ in the MSSM}
\begin{figure}[t]
    \begin{center}
       \setlength{\unitlength}{1truecm}
       \begin{picture}(10.0, 9.0)
       \epsfxsize 10.  true cm
       \epsffile{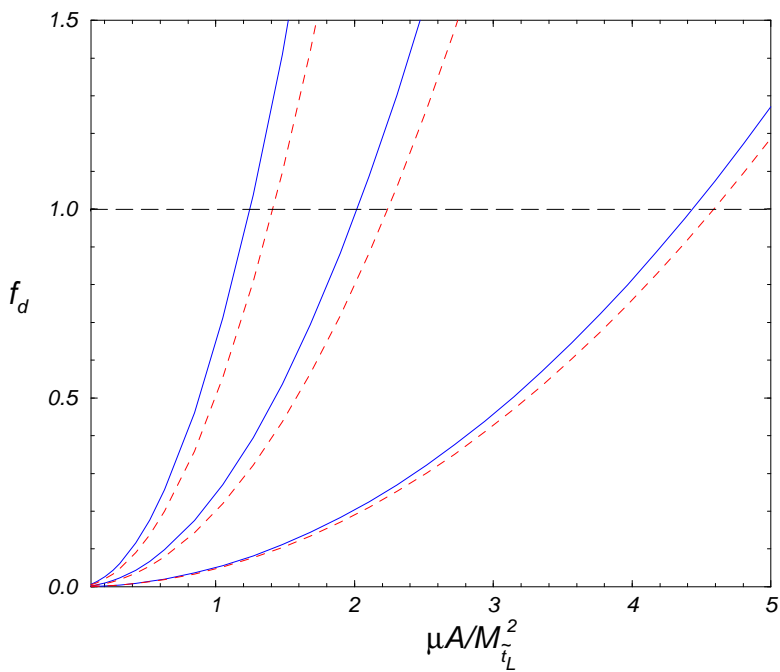}
       \end{picture}
    \end{center}
    \caption{$f_d$ as a function of $\mu A/M^2_{{\tilde{t}}_L}$ for 
$\mu=100$ GeV and different stop masses. The three solid (dashed) 
curves from the left are obtained for $\{ M_{{\tilde{t}}_{R}}({\rm GeV}), 
   ~M_{{\tilde{t}}_{L}}({\rm GeV}) \} = \{ 200,~500~(300) \}$, $\{300,~1000~(500)\}$, 
  $\{ 500,~1000~(800)\}$.}
    \protect\label{fig:f}
\end{figure}

In the case of $\Delta M_{d}$, the only relevant
$\Delta B=2$ scalar operator is $Q^{SLL}$. Neglecting possible 
new-physics effects in the standard vector-current operator,
according to Eqs.~(\ref{eq:gs1}) and (\ref{eq:fds})
the large-$\tan\beta$ contribution to $f_d$ is given by 
\bea
f_{d}&=&\frac{{\bar P}^{SLL} C^{SLL}(\mu_t)}{S_0(x_{tW})} 
    \approx \frac{{ 4\bar P}^{SLL}}{S_0(x_{tW})} 
    \frac{m^4_t m^2_b \tan^2 \beta}{M^2_W \mu^4}
    \left(\frac{\mu A}{M^2_{{\tilde{t}}_L}}\right)^2   
      \frac{G(x_{R \mu},x_{L \mu})}{(1-x_{R L})^2}\nnu~.
\eea
Imposing that $f_d$ varies within the range allowed by the 
measurement of $\Delta M_{d}$, we can derive a set 
of bounds on the quantity $\mu A/M^2_{{\tilde{t}}_L}$ 
for different values of stop masses ($M_{{\tilde{t}}_L}$
and $M_{{\tilde{t}}_R}$) and $\mu$, as shown in Fig.~\ref{fig:f}.
Although the experimental determination of $\Delta M_{d}$ is rather precise, 
once the uncertainties  on ${\hat B}_{B_{d}} F^2_{B_{d}}$ 
and $|V_{td}|$ are taken into account, $f_d$ can reach
values close to 1, which we take as a reference figure.
As can be noted from  Fig.~\ref{fig:f}, the bounds on 
 $\mu A/m^2_{{\tilde{t}}_L}$ thus obtained are not severe.
Even for small $\mu$
($\mu\approx 100$ GeV), light $m_{{\tilde{t}}_R}$ 
($m_{{\tilde{t}}_R}\approx 200$ GeV) 
and a large stop splitting, the upper bound on 
$\mu A/m^2_{{\tilde{t}}_L}$ is above unity, 
condition that is naturally satisfied within our scenario and 
that still allows large effects in $\Delta B=1$ processes.

Constraints even less strict on $\mu A/M^2_{{\tilde{t}}_L}$
can in general be set by using the present experimental lower 
bound for the $B_s$--${\bar B}_s$ mixing. 
In this case also the double-penguin contribution to $Q^{SLR}$
has to be taken into account. As shown in Eq.~(\ref{eq:LLvLR}), for 
light $\mu$ and $M_{A} \sim \mu$, the large $\tan\beta$
contributions to $C^{SLR}$ and $C^{SLL}$ are comparable
and tend to cancel out in $f_s$, owing to the opposite sign 
between ${\bar P}^{SLL}$ and ${\bar P}^{SLR}$.
Only if $M_{A}$ is light and $\mu \gg M_{A}$, or when the 
double-penguin contribution completely dominates, does one obtain
a non-trivial bound. 

In the limit where the double-penguin contribution dominates,
higher-order $\tan\beta$-enhanced corrections are numerically
relevant and cannot be ignored. In particular, terms 
proportional to \cite{HRS}
\beq
\epsilon_0 = \frac{ 2 \alpha_S}{3\pi} \frac{\mu M_3 }{M_{\widetilde d_L}^2  } 
f\left( x_{3 L} , x_{R L} \right)~,
\label{eq:eps0_susy}
\eeq
turn out to be very large. In order to obtain a simple expression, 
we can neglect the contribution of $\epsilon_Y y_t^2  $ with 
respect to $\epsilon_0$ in the denominator of Eq.~(\ref{eq:CSLRd}),
and set $\epsilon_0 \approx 1/100$, as obtained in the case 
of degenerate sparticles. We also set $\epsilon_0>0$, following 
the indication $\mu>0$ from $b\to s\gamma$ and $(g-2)_\mu$ \cite{g_2_mw}.  
We then obtain 
\beq
f_{s} = \frac{{\bar P}^{SLR} C^{SLR}(\mu_t)}{S_0(x_{tW})} 
    \approx - 0.4 \left[ \frac{ 3(\tan\beta/50) }{2+(\tan\beta/50)}  \right]^4  
            \left( \frac{\rm 200~GeV}{M_A} \right)^2 
    \left(\frac{\mu A f(x_{\mu L},x_{R L}) }{M_{\widetilde t_L}^2  } \right)^2~.
\label{eq:fd_2p}
\eeq
Since the present lower bound on $\Delta M_s$ implies a lower bound
on $f_{s}$ of about $-0.5$ \cite{Buras}, we conclude that also 
in this scenario values of $\mu A/M^2_{{\tilde{t}}_L} \lsim 1$ 
are still allowed. Note that a much stronger bound would have been 
obtained without the inclusion of higher-order terms: for $\tan\beta \approx 50$
these reduce the lowest-order result almost by a factor 5.

\section{Phenomenology of $B_{s,d} \to \ell^+ \ell^-$ decays}

\subsection{Effective Hamiltonian}
Within the SM there is only one dimension-six operator 
generating a non-vanishing contribution to $B_q \to \ell^+\ell^-$
decays, namely 
\beq
Q_{10}  = \bar{b}_L \gamma^\mu q_L \bar{\ell} \gamma_\mu \gamma_5 \ell~.
\label{eq:q10}
\eeq
Being scale-invariant and completely dominated 
by short-distance dynamics, its Wilson coefficient 
is well approximated by the lowest-order result \cite{BBL}:
\beq
\left. C^{\rm SM}_{10}\right|_{\rm LO} 
=  \frac{G_F^2 M_W^2 V^*_{tq}  }{\pi^2}~ Y_0(x_{tW}) \approx 
  \frac{G_F^2 M_W^2 V^*_{tq}}{\pi^2}~ \left(\frac{m_t(\mu_t)}{\rm 166~GeV}\right)^{1.56}~.
\eeq
The full NLO expression of $C_{10}^{\rm SM}$ can be found in \cite{MU}. 

The additional scalar interactions that appear
in the SM extensions under investigation 
are fully described by the operators in Eq.~(\ref{eq:sc_ops}).
If the lepton Yukawa coupling is not negligible,
as in the case of a $\tau^+\tau^-$ final state,  in principle 
we should take into account also the right-handed vector-current operator $Q_{10}'$, 
obtained by $Q_{10}$ under the exchange $(q,b)_L \to (q,b)_R$.
However, as can be expected from a naive extrapolation of 
the results in the previous section, the numerical impact of this operator
is always negligible. Within the 2HDM one finds
\bea
C'^H_{10} = \frac{G_F^2 M^2_W V^*_{tq} }{\pi^2} 
      \frac{m_q m_b m^2_\ell }{4 M^2_W M^2_H} \tan^4\beta B_0(x_{tH})~,
\eea
where $B_0(x)=[L_3(x,0,1)-L_3(0,0,1)]/4$. In the  
most favourable case, namely the $B_s \to \tau^+ \tau^-$ decay,
$C'^H_{10}$ leads at most to $\cO(1\%)$ corrections  
to the SM result. A similar suppression is found  
also for the MSSM contribution to $C'_{10}$.

For completeness, we report here also the MSSM chargino-box
contribution to $C_{S,P}$, which was not explicitly 
shown in section~\ref{sec:DF=1}. This is given by 
\beqa
C^{\chi}_{S,P} &=& \frac{G_F^2 M^2_W}{2\pi^2} V_{3i}^* V_{3j} 
    \sum_{h,k,m,n}   x_{W \chi_n}
    [  Z^b_{m {\tilde t}_h} Z^\ell_{m {\tilde \nu}_k} Y_{n {\tilde \nu}_k} 
        Y_{n {\tilde t}_h}  
        L_3(x_{{\tilde \nu}_k \chi_n},x_{{\tilde t}_h \chi_n},
        x_{\chi_m\chi_n })\nnu\\ 
        &\pm& \frac{m_{\chi_m}}{m_{\chi_n}} 
         Y_{m {\tilde t}_h} Z^\ell_{m {\tilde \nu}_k} Y_{n {\tilde \nu}_k} 
        Z^b_{n {\tilde t}_h}  
        L_4(x_{{\tilde \nu}_k \chi_n},x_{{\tilde t}_h \chi_n},
        x_{\chi_m\chi_n })] 
        - ({\tilde t}\rightarrow {\tilde u})~,   \label{eq:box_al}
\eea
where we have assumed diagonal chargino--lepton--slepton couplings,
and terms suppressed by $m_{s,d}/m_b$ or by the 
mass difference between the first two generations of squarks 
have been neglected (under these hypotheses Eq.~(\ref{eq:box_al}) 
is agreement with the results of Ref.~\cite{Bobeth}). 
Following the simplifying assumptions made
for the chargino--squark contributions to $B^0$--$\bar B^0$ mixing, 
the above result becomes
\bea
C^{\chi}_S = C^{\chi}_P &=& 
     \frac{G_F^2 M^2_W}{4\pi^2} V_{3i}^* V_{3j}
     \frac{m_b m_\ell \tan^2 \beta}{\mu^2}  
      \left[ \cos^2\theta_t 
       L_3(x_{{\tilde \nu} \mu},x_{{\tilde t}_1 \mu},x_{{\tilde W}\mu}) \right.  \no\\ 
      && \left. \qquad\qquad   + \sin^2\theta_t L_3(x_{{\tilde \nu} \mu},
       x_{{\tilde t}_2 \mu},x_{{\tilde W}\mu})
       - L_3(x_{{\tilde \nu} \mu},x_{{\tilde u} \mu},x_{{\tilde W}\mu}) \right].
\eea
From this expression it is straightforward to check that,  
for large $\tan \beta$, box contributions cannot compete 
with the neutral Higgs-penguin coefficients 
in Eq.~(\ref{eq:susy2}), as already 
anticipated in section~\ref{sec:DF=1}.
We shall therefore neglect such terms in the following 
phenomenological discussion.

\subsection{Branching ratios}
Writing hadronic matrix elements of axial and pseudoscalar 
currents as 
\beq
\langle\bar 0 |\bar q \gamma_\mu\gamma_5  b|\bar B_q(p)\rangle = i p_\mu f_{B_q}~, 
\qquad \langle\bar 0 |\bar q \gamma_5  b|\bar B_q(p)\rangle =  -i f_{B_q} 
\frac{M^2_{B_q } }{ (m_b+m_q )}~, 
\label{eq:fbdef}
\eeq
the most general expression of $B_q \to \ell^+ \ell^-$ 
branching ratios reads, according to our normalization of 
the effective operators:
\bea
{\mathcal B}\,(B_q \to \ell^+ \ell^-) 
   &=& \frac{  f^2_{B_q } M_{B_q } \tau_{B_q } m^2_\ell }{8\pi}
    \left(1-\frac{4 m^2_\ell}{M^2_{B_q }}\right)^{1/2} \left[ 
       \left(1-\frac{4 m^2_\ell}{M^2_{B_q }}\right) 
       \left| \frac{M^2_{B_q } ( C_S-C'_S)  }{2 m_\ell (m_b+m_q ) } \right|^2 \right. \no\\
    &&\left.  +\left| \frac{M^2_{B_q}(C_P-{C'}_P)}{2m_\ell (m_b+m_q) }
       + C_{10}-{C'}_{10} \right|^2 ~ \right]~,
\eea
where light-quark masses and Wilson coefficients are understood 
to be evaluated at a scale of $\cO(m_b)$. 
Since the ratios $(C_{S,P}-C'_{S,P})/m_\ell$ are independent of $m_\ell$,
to a first approximation the relative weight of the various contributions is 
independent of $m_\ell$, whereas the overall branching ratio scales 
like $m_\ell^2$. This leads to a strong enhancement factor
of the $\tau^+\tau^-$ modes that partially compensate their difficult detection. 

Within the SM, employing the full NLO expression of
$C_{10}^{\rm SM}$ \cite{MU}, the branching ratios 
of the two $B_s$ modes are given by 
\beqa
\left. {\mathcal B}( B_s \to \mu^+ \mu^-)\right|_{\rm SM} &=& 3.1 \times 10^{-9}
\left( \frac{f_{B_s}}{0.21~\mbox{GeV}} \right)^2
\left( \frac{|V_{ts}|}{0.04}           \right)^2
\left( \frac{\tau_{B_s}}{1.6~\mbox{ps}} \right)
\left( \frac{ m_t(m_t) }{166~\mbox{GeV} } \right)^{3.12} \quad
\label{eq:BrmmSM} \\
\left. {\mathcal B}( B_s \to \tau^+ \tau^-)\right|_{\rm SM} &=& 215 \times 
\left. {\mathcal B}( B_s \to \mu^+ \mu^-)\right|_{\rm SM}~,
\label{eq:BrttSM} 
\eeqa
whereas the corresponding $B_d$ modes are both 
suppressed by an additional factor $|V_{td}/V_{ts}|^2$ 
$=(4.0 \pm 0.8)\times10^{-2}$. Note that the 
${\mathcal B}\,(B_{s}\to \tau^+ \tau^-)/{\mathcal B}\,(B_{s}\to \mu^+ \mu^-)$
ratio in Eq.~(\ref{eq:BrttSM}) can be considered as a model-independent  
upper bound: if $C_{S,P} \propto m_\ell$ this ratio can only be smaller 
than in the SM case. At present
the experimental bound closest to SM expectations is given by
\be
{\mathcal B}\,(B_{s}\to \mu^+ \mu^-) < 2.6 \times 10^{-6} \;\;
(95 \% \;\;  {\rm CL}) \;\; \cite{CDF}~.
\label{eq:CDF}
\ee

Working in the limit where the supersymmetric contribution to 
$C_{S,P}$ in Eq.~(\ref{eq:susy2}) dominates over all other 
contributions, and neglecting terms suppressed by $m_s/m_b$ and $m_\mu/m_b$, 
we find
\beqa
\left. {\mathcal B}\,(B_s \to \mu^+ \mu^-)\right|_{\rm MSSM}
  &=& \frac{  f^2_{B_s } G_F^4 M^5_{B_s } |V_{ts}|^2 m^2_\mu \tau_{B_s }  }{256\pi^5}
%%  \left( \frac{  m_b(\mu_t)}{ m_b(m_b) } \right)^2 
  \left( \frac{m_t(\mu_t) }{  M_A } \right)^4  
  \left(\frac{\mu A}{M_{\widetilde t_L}^2  } \right)^2 
  \no \\ && \qquad
 \times \left( \frac{ \tan^3\beta ~ f(x_{\mu L},x_{R L}) }{ (1+\epsilon_0 \tan \beta) 
            [  1+ \tan \beta ( \epsilon_0  +  \epsilon_Y y_t^2 ) ] } \right)^2~.
\label{eq:BmmSUSY}
\eeqa
Setting $\epsilon_0 \approx 1/100$ and neglecting the $\epsilon_Y y_t^2$ term
in the denominator [as done in Eq.~(\ref{eq:fd_2p})], and employing the reference 
values of $f_{B_s}$, $\tau_{B_s }$, $m_t$ and $V_{ts}$ as in Eq.~(\ref{eq:BrmmSM}),
we obtain the following compact phenomenological expression 
\beq
\left. {\mathcal B}\,(B_s \to \mu^+ \mu^-)\right|_{\rm MSSM} \approx 
  3 \times 10^{-6}  
 \frac{ r^6 }{ \left( \frac{2}{3} + \frac{1}{3}r \right)^4 }
 \left( \frac{200~\rm{GeV}}{M_A} \right)^4  
 \left(\frac{\mu A  f(x_{\mu L},x_{R L}) }{ M_{\widetilde t_L}^2  } \right)^2 ~,
\label{eq:BmmSUSY2}
\eeq
where $r=\tan\beta/50$. Given the weak constraints on 
$\mu A/M_{\widetilde t_L}^2$ discussed in section~\ref{sec:BB}, 
we cannot exclude a scenario where ${\mathcal B}\,(B_s \to \mu^+ \mu^-)$
is just below its present experimental bound, as already pointed 
out in Refs.~\cite{Babu,Chanko,Bobeth}. However, we stress that this can occur 
only for rather extreme values of the parameter space, in particular 
for $\tan\beta \sim 50$, when higher-order $\tan\beta$-enhanced 
corrections are very large. For $\tan\beta \leq 30$ it is 
already unlikely to find ${\mathcal B}\,(B_s \to \mu^+ \mu^-)$ above the 
$10^{-7}$ level and for $\tan\beta \leq 10$ supersymmetric 
contributions do not exceed the level of the SM ones.
Concerning higher-order $\tan\beta$-enhanced terms, we 
recall that these are fully taken into account 
by means of Eq.~(\ref{eq:BmmSUSY}), once the appropriate expressions of 
$\epsilon_0$ and $\epsilon_Y$ are used, whereas
the approximate result in Eq.~(\ref{eq:BmmSUSY}) is valid only
for $M_3 \sim M_{\tilde q} \sim \mu$ ($\mu >0$).

In the scenario where the present discrepancy between 
experimental data and SM predictions on the anomalous magnetic
moment of the muon \cite{g_2_exp,g_2_cm} is due to supersymmetric effects, 
the possibility of a sizeable enhancement of ${\mathcal B}\,(B_s \to \mu^+ \mu^-)$
certainly becomes more likely. Indeed in both cases supersymmetric 
contributions grow with $\tan\beta$. As recently pointed out in 
Ref.~\cite{Nierste}, the correlation between these two observables 
becomes very strong in the constrained minimal supergravity scenario.
Unfortunately, the situation is not so clear in the general framework
of the MSSM. A useful tool to illustrate basic features of 
the supersymmetric contribution to $a_\mu=(g-2)_\mu/2$ is the expression
\beq
\frac{a^{\rm MSSM}_\mu}{ 15 \times 10^{-10}}  \approx
1.7  \left(\frac{\tan\beta }{50} \right) 
\left( \frac{500~\rm GeV}{M_{\tilde \chi}} \right)^2
\left( \frac{M_{\tilde \nu}}{M_{\tilde \chi}} \right)~,
\label{eq:g_2}
\eeq
which provides a good approximation to the full
one-loop result \cite{g_2_mw} in the limit of almost degenerate higgsinos and 
electroweak gauginos ($M_1 \sim M_2 \sim \mu \gg M_W$),
and allowing a moderate splitting between slepton and chargino masses
(the normalization of the l.h.s is the 
present experimental error \cite{g_2_exp}).
Comparing Eqs.~(\ref{eq:BmmSUSY2}) and (\ref{eq:g_2}) helps us 
draw the following general conclusions, whose validity 
goes beyond the approximations made to derive the two equations:
\begin{itemize}
\item{} The mild $\tan\beta$ dependence of $a_\mu^{\rm MSSM}$
allows significant supersymmetric contributions to the latter, 
even for $\tan\beta \leq 10$. In this case 
the effects on ${\mathcal B}\,(B_s \to \mu^+ \mu^-)^{\rm }$ would
be undetectable, at least before the LHC.
\item{} ${\mathcal B}\,(B_s \to \mu^+ \mu^-)^{\rm MSSM}$
strongly depends on $A$ and $M_A$, which play almost no role
in $a_\mu^{\rm MSSM}$. As a result, 
${\mathcal B}(B_s \to \mu^+ \mu^-)^{\rm MSSM}$  can be made arbitrarily 
small even if $a^{\rm MSSM}_\mu$ saturates the experimental result
and $\tan\beta \sim 50$.
\item{} A detection of ${\mathcal B}(B_s \to \mu^+ \mu^-)$
at the $10^{-7}$ level (or above) would provide a strong support 
in favour of the supersymmetric explanation of the $(g-2)_\mu$ puzzle. This result 
could be interpreted in the MSSM framework only for $\tan\beta \gsim 30$ 
and light $M_A$, a scenario where it is difficult to 
keep $a_\mu^{\rm MSSM}$ below $10^{-9}$ 
unless chargino masses are unnaturally high.
\end{itemize}

\medskip

Concerning $B_{s,d} \to \tau^+\tau^-$ modes, the following relations hold in 
the limit where scalar-current contributions are dominant:
\beqa
\frac{{\mathcal B}\,(B_{s}\to \tau^+ \tau^-)}{{\mathcal B}\,
     (B_{s}\to \mu^+ \mu^-)} &=& \frac{m^2_\tau}{m^2_\mu} 
    \sqrt{ 1-\frac{4m_\tau^2}{M_{B_{s}}^2}}
    \left(1-\frac{2m_\tau^2}{M_{B_{s}}^2}\right) = 166~, \label{eq:r1} \\
\frac{{\mathcal B}\,(B_{d}\to \tau^+ \tau^-)}{{\mathcal B}\,
     (B_{s}\to \mu^+ \mu^-)} &=& \frac{ f^2_{B_d}M^5_{B_d}|V_{td}|^2\tau_{B_d}
     }{  f^2_{B_s}M^5_{B_s}|V_{ts}|^2\tau_{B_s} } \frac{m^2_\tau}{m^2_\mu} 
    \sqrt{ 1-\frac{4m_\tau^2}{M_{B_{d}}^2}}
    \left(1-\frac{2m_\tau^2}{M_{B_{d}}^2}\right) = 6.8 \pm 1.4~.
\label{eq:Bdtt}
\eeqa
By means of Eq.~(\ref{eq:CDF}), these allow us to set the following indirects limits:
\be
{\mathcal B}\,(B_{s}\to \tau^+ \tau^-) < 4.3 \times 10^{-4} \; , \qquad
{\mathcal B}\,(B_{d}\to \tau^+ \tau^-) \lsim 2 \times 10^{-5} \; .
\ee
Unfortunately the present experimental information on the $\tau^+\tau^-$ modes is 
very poor: we are not aware of any progress with respect to the indirect bounds, 
at the per cent level, discussed in Ref.~\cite{Yuval}. The signature of 
these modes is certainly very difficult at hadron colliders, 
but significant improvements could be expected from $B$ factories. 
Reaching a $10^{-5}$ sensitivity on the  $B_{d}\to \tau^+ \tau^-$ mode, $B$-factory 
experiments could compete with the present search for 
$B_{s}\to \mu^+ \mu^-$ at Fermilab.

\section{Conclusions}
In this paper we have presented a general analysis of 
scalar FCNC amplitudes of down-type quarks in 
two classes of models with two Higgs doublets: the MSSM 
and the non-supersymmetric 2HDM of type II. 
We have shown that the $\tan\beta$ growth 
of these amplitudes is not simply determined by the number 
of Yukawa couplings appearing in the amplitudes. A crucial role 
is also played by the breaking of the $U(1)_d$ symmetry 
that forbids the tree-level coupling of $H_u$
to down-type quarks. Within the MSSM, the $U(1)_d$
breaking induced by the ${\widetilde H}_u$--${\widetilde H}_d$ mixing
is not necessarily suppressed in the large-$\tan\beta$
limit, resulting in a potential
$\tan^2\beta$ growth of the effective $\bar{d}_R^k d_L^j H_d$ vertex 
generated at one loop. This {\em anomalous} $\tan\beta$
behaviour is characteristic of the
effective FCNC coupling of down-type quarks to neutral Higgses
and it does not appear in irreducible $\Delta F=2$ amplitudes
at the one-loop level. On the other hand, because of the $1/\tan\beta$ suppression 
of the ordinary $H_u$--$H_d$ mixing, this phenomenon does not arise 
at all in the non-supersymmetric model.
As explicitly shown, in both cases leading $\tan\beta$ 
contributions to scalar amplitudes can efficiently  be
computed in the gaugeless limit of the models, or considering 
only the Higgs sector of the theory. 

Despite the potential large enhancements (if $\tan\beta \gsim m_t/m_b$), 
non-standard scalar FCNC amplitudes are difficult to be identified 
experimentally. The rare dilepton decays $B_{s,d} \to \ell^+\ell^-$
offer an almost unique opportunity in this respect.
As pointed out first in Ref.~\cite{Babu}, the $\tan^3\beta$
growth of the $b \to s \mu^+ \mu^-$ scalar amplitude, within the MSSM, 
could lead to order-of-magnitude enhancements of  
the $B_s \to \mu^+ \mu^-$ rate compared to SM expectations. 
We have explicitly 
checked that this statement remains true even when taking into 
account the existing constraints on $\Delta B=2$ scalar 
operators. The latter includes two types of effects: i) the reducible two-loop 
contribution proportional to $y_b y_{s,d} \tan^4 \beta$ discussed in 
Ref.~\cite{Buras}; ii) an irreducible one-loop contribution proportional 
to $y_b^2 \tan^2 \beta$. The two effects (comparable in size 
in the $B_s$--${\bar B_s}$ case) are typically smaller than the SM 
amplitude and weakly constrained at present.

The order of magnitude enhancements of  $B_{s,d}\to \ell^+ \ell^-$
rates can occur only for $\tan\beta \gsim 30$, when higher-order 
$\tan\beta$-enhanced terms cannot be neglected.
Following the approach of Ref.~\cite{bsgamma}, 
we have discussed how to control these terms beyond 
lowest order. Despite a drastic reduction of the  
one-loop result due to higher-order $\tan\beta$-enhanced  terms 
(up to a factor 5 for $\tan\beta \sim 50$), the possibility that 
${\mathcal B}\, (B_s \to \mu^+ \mu^-)$ is just below its 
present experimental bound is still open. 
An evidence of ${\mathcal B}\, (B_s \to \mu^+ \mu^-)$ 
at the $10^{-7}$ level (or above) would provide a strong support 
in favour of a supersymmetric explanation of the $(g-2)_\mu$ puzzle \cite{Nierste};
however, we have shown that 
the opposite is not true in the general MSSM framework: a sizeable 
supersymmetric contribution to $(g-2)_\mu$ does not necessarily 
imply observable consequences in  ${\mathcal B}\, (B_s \to \mu^+ \mu^-)$.
Finally, we have emphasized the interest of $B_{s,d} \to \tau^+\tau^-$ modes
in probing non-standard scalar FCNC amplitudes and, particularly, the 
importance to search for $B_d \to \tau^+\tau^-$ at $B$ factories.

\section*{Acknowledgements}
We thank A.J. Buras and P. Gambino for useful discussions.

\newpage 

\appendix
\section{Loop functions}
\setcounter{equation}{0}

\bea
S_0(x) &=& \frac{4x-11x^2+x^3}{4(1-x)^2} - \frac{3}{2} \frac{x^3 \log{x}}{(1-x)^3}\; , 
           \qquad\qquad
           S_0\left( \frac{m_t^2}{M^2_W} \right) = 2.4 \pm  0.2 \; , \no\\
f(x,y) &=& \frac{1}{y-x}\left[\frac{x \log x}{1-x}-\frac{y \log y}{1-y}\right]\; , 
           \qquad\qquad\quad\
           f(1,1) = \frac{1}{2} \; ,   \no\\
F(x,y,z)   &=& \frac{x \log{x}}{(x-1)(x-y)(x-z)} \; , \no\\
L_1(x,y,z) &=& xy[F(x,y,z)+F(y,z,x)+F(z,x,y)] \; ,      
           \qquad\qquad  L_1(1,1,1) = -\frac{1}{6} \; ,  \qquad      \no\\ 
L_2(x,y,z) &=& xy[xF(x,y,z)+yF(y,z,x)+zF(z,x,y)] \; ,  
           \qquad\  L_2(1,1,1) = \frac{1}{3} \; ,         \no\\
L_3(x,y,z) &=& \frac{1}{xy} L_2(x,y,z) \; ,     \no\\  
L_4(x,y,z) &=& \frac{1}{xy} L_1(x,y,z) \; ,     \no\\ 
G_2(y)&=& \frac{1 - 9y - 9y^2 + 17y^3 - 6y^2(3 + y) \log y}{6(y-1)^5}~,
           \qquad\quad\quad\!\!  G_2(1) = -\frac{1}{20} \; ,   \no\\
\no
\eea

%\footnotesize

\end{document}